\documentclass[%
 reprint,
superscriptaddress,
 amsmath,amssymb,
aps,
prb,
floatfix,
]{revtex4-2}
\usepackage{silence}
\usepackage{lmodern}
\usepackage{siunitx}
\usepackage{chemfig}
\usepackage{amsmath}
\usepackage{amssymb}
\usepackage{color}
\usepackage{times}
\usepackage{epsf}
\usepackage{graphicx}% Include figure files
\usepackage{epsfig}
\usepackage{epstopdf}
\usepackage{dcolumn}% Align table columns on decimal point
\usepackage{bm}% bold math
\usepackage{float}
\usepackage{url}
\usepackage{CJK}
\usepackage{ulem}
\usepackage{makecell}
\usepackage{newtxtext, newtxmath}
\usepackage{multirow}
\usepackage{subfigure}
\usepackage{braket}
\usepackage[colorlinks=true,linkcolor=blue,citecolor=blue,urlcolor=blue]{hyperref}

\def\<{\langle}
\def\>{\rangle}
\def\({\left(}
\def\){\right)}
\def\[{\left[}
\def\]{\right]}

\begin{document}
%\title{Crystalline Pseudoangular Momentum Governs Selective Magnon–Phonon Hybridization}
\title{{Equivalent Pseudo-angular Momentum Governs Selective Magnon--Phonon Hybridization}}
\author{Hong Sun}
\affiliation{Ministry of Education Key Laboratory of NSLSCS, Phonon Engineering Research Center of Jiangsu Province, Center for Quantum Transport and Thermal Energy Science, Institute of Physics Frontiers and Interdisciplinary Sciences, School of Physics and Technology, Nanjing Normal University, Nanjing 210023, China}

\author{Hao Chen}
\email{phych@njupt.edu.cn}
\affiliation{School of Science, Nanjing University of Posts and Telecommunications, Nanjing, 210023, China}

\author{Lifa Zhang}
\email{phyzlf@njnu.edu.cn}
\affiliation{Ministry of Education Key Laboratory of NSLSCS, Phonon Engineering Research Center of Jiangsu Province, Center for Quantum Transport and Thermal Energy Science, Institute of Physics Frontiers and Interdisciplinary Sciences, School of Physics and Technology, Nanjing Normal University, Nanjing 210023, China}

\date{\today}
\begin{abstract}
{Magnon--phonon hybridization provides a powerful route to couple spin and lattice excitations, yet the symmetry principle governing mode selectivity remains unclear. Here, we show that crystalline pseudoangular momentum provides a general criterion for selective magnon--phonon hybridization in noncollinear multisublattice magnetic systems. The rotation eigenvalue associated with pseudoangular momentum encodes polarization transformations, sublattice permutations, and Bloch phases, requiring magnetoelastic coupling between modes belonging to the same rotation sector. We illustrate this principle in a stacked kagome model with noncollinear $120^\circ$ spin order. An out-of-plane magnetic field tunes one magnon sector into resonance with a degenerate phonon doublet, yet only the pseudoangular-momentum-matched phonon develops a pronounced avoided crossing while the other remains nearly unmixed. Field reversal switches the selected rotation sector and reverses the rotational polarization of the hybridized phonon component without changing the hybrid-mode energies. Our results reveal pseudoangular-momentum matching as a general framework for understanding and controlling mode-selective spin--lattice hybridization.}
%We establish crystalline pseudoangular momentum as a symmetry criterion for selective magnon--phonon hybridization in noncollinear multisublattice magnetic systems. Because the corresponding rotation eigenvalue encodes polarization transformations, sublattice permutations, and Bloch phases, a rotation-symmetric magnetoelastic interaction couples only magnon and phonon modes in the same rotation sector. We demonstrate this selection rule in a stacked kagome model with noncollinear $120^\circ$ order. An out-of-plane field tunes one magnon into resonance with two degenerate phonons, yet only one forms a pronounced avoided crossing while the other remains predominantly phononic. Reversing the field switches the selected rotation sector and the phonon rotational polarization without changing the hybrid-mode energies. Our results distinguish symmetry compatibility from energetic resonance and provide a route to field-controlled mode-selective spin--lattice hybridization.
\end{abstract}
\maketitle
\textit{Introduction.}\textemdash Magnons and phonons are elementary collective excitations of magnetic solids, describing spin precession and lattice vibrations, respectively. When their dispersions meet at the same crystal momentum, spin--lattice interactions can hybridize these modes, giving rise to an avoided crossing and hybrid quasiparticles known as magnon polarons~\cite{kittel1958interaction,PhysRevLett.117.207203,bao2023direct}. Near resonance, these quasiparticles carry both spin and lattice character, with the associated avoided crossings generating Berry-curvature hot spots~\cite{PhysRevLett.117.217205,PhysRevB.99.174435,PhysRevLett.123.167202,PhysRevLett.123.237207} and strongly reshaping thermal and spin transport~\cite{PhysRevB.95.144420,PhysRevB.96.104441,PhysRevLett.127.247202}. Magnon--phonon hybridization is now a prominent theme in spin--lattice physics.
\par
Recently, a new question has emerged concerning this hybridization: when degenerate magnon and phonon multiplets overlap in energy, why do only selected mode pairs hybridize? A striking example is the experimental observation of chirality-selective coupling in the layered antiferromagnet FePSe$_3$, where only selected pairs of nearly degenerate magnons and optical phonons hybridize even though other pairs lie comparably close in energy~\cite{cui2023chirality}. This selectivity was attributed to angular-momentum matching between spin precession and atomic rotational motion, suggesting that phonon chirality provides an additional mode-selection criterion beyond spectral resonance. Subsequent theoretical studies showed that the circular polarization~\cite{PhysRevLett.115.115502} and associated angular momentum~\cite{PhysRevLett.112.085503} of phonons can govern selective magnon--phonon coupling at the $\mathrm{K}$ and $\mathrm{K}^{\prime}$ valleys of hexagonal lattices~\cite{PhysRevB.108.174426}, that symmetry-allowed magnetoelastic anisotropy can generate chiral phonons~\cite{PhysRevLett.133.246604}, and that chirality-selective magnon coupling to transverse phonons can arise even in inversion-symmetric ferromagnets~\cite{j7bs-2zbx}. More recently, chirality-selective magnon–phonon hybridization was predicted to generate a $d$-wave phonon-angular-momentum texture in a collinear altermagnet~\cite{r43h-ch8r}. Taken together, these studies establish phonon chirality and angular momentum as key selectors of spin--lattice hybridization.
\par
However, from a broader perspective, local handedness alone does not provide a complete criterion for such selectivity. In multisublattice crystals, magnetoelastic coupling depends on the full spatial structure of the participating modes; it is their relative phase pattern---rather than the local chirality or angular momentum of either mode alone---that determines whether symmetry-related contributions reinforce or cancel one another. In the collinear settings considered in these studies, spin precession and ionic rotation share a common reference axis, making their relative handedness a natural proxy for phase compatibility. In a noncollinear multisublattice magnet, however, this correspondence breaks down: no single axis characterizes all local spin precessions, while at finite crystal momentum a crystal rotation also involves sublattice permutations and the accompanying Bloch phases. Local circular polarization, physical angular momentum, and crystalline pseudoangular momentum therefore need not coincide. It thus remains unclear what determines the mode selectivity of magnon--phonon hybridization in such systems.
\par
In this Letter, we show that crystalline pseudoangular momentum provides a general criterion. At a crystal momentum invariant under an $n$-fold rotation up to a reciprocal-lattice vector, the bare magnon and phonon modes can be resolved into eigenstates of the corresponding little-group rotation operator of the magnetically ordered state. If the magnetoelastic interaction preserves this rotational symmetry, a magnon and a phonon can mix only if their rotation eigenvalues match, yielding the selection rule $l_{\mathrm{m}}=l_{\mathrm{ph}}\pmod{n}$. Unlike labels based only on local handedness or mode angular momentum, this rotation eigenvalue incorporates the transformation of the local spin or displacement polarization together with the sublattice permutation and accompanying Bloch phases. It thereby characterizes the collective symmetry compatibility between magnon and phonon modes that underlies their selective hybridization.
\par
We demonstrate this principle in a stacked kagome-lattice model with a $120^\circ$ noncollinear spin texture within each layer and antiferromagnetic alignment between adjacent layers. The displacement dependence of the Dzyaloshinskii--Moriya (DM) interaction generates a linear coupling between magnons and in-plane phonons. At the $\mathrm{K}$ point, the relevant magnon and phonon modes each form a twofold-degenerate pair in the absence of an applied magnetic field. Under $C_3$, both doublets decompose into $l=0$ and $l=1$ modes; however, the two symmetry-allowed coupling channels are equivalent, so there is no observable mode selectivity. An out-of-plane magnetic field preserves ${C}_3$ but lifts the magnon degeneracy while leaving the phonons degenerate, bringing one magnon sector into resonance with both phonons. Yet this magnon develops a pronounced avoided crossing only with the phonon of matching pseudoangular momentum, while the other remains nearly unmixed. Reversing the field switches the resonant magnon sector and, in the present model, reverses the rotational polarization of the hybridized phonon component. These results show that, beyond energetic resonance, pseudoangular-momentum matching also provides a general symmetry criterion for mode-selective magnon--phonon hybridization in noncollinear magnets.
\par
\textit{Selection rule.}\textemdash To make it explicit, we consider a magnetically ordered system that preserves an $n$-fold rotation $C_n$. For a wave vector $\mathbf{k}$ satisfying ${C}_n\mathbf{k}=\mathbf{k}+\mathbf{G}$, where $\mathbf{G}$ is a reciprocal-lattice vector, ${C}_n$ belongs to the little group of $\mathbf{k}$. The uncoupled magnon and phonon Hamiltonians at such a wave vector therefore commute with the corresponding rotation operator $\hat{C}_n$, allowing their eigenmodes to be chosen with definite rotation eigenvalues,
\begin{equation}\label{eq1}
\begin{aligned}
\hat{C}_{n}
\ket{\mathrm{m}_{\mu\mathbf{k}}}&
=e^{-i2\pi l_{\mu}^{\mathrm{m}}/n}
\ket{\mathrm{m}_{\mu\mathbf{k}}},\\
\hat{C}_{n}
\ket{\mathrm{ph}_{\nu\mathbf{k}}}&
=e^{-i2\pi l_{\nu}^{\mathrm{ph}}/n}
\ket{\mathrm{ph}_{\nu\mathbf{k}}},
\end{aligned}
\end{equation}
where $l_{\mu}^{\mathrm{m}}$ and $l_{\nu}^{\mathrm{ph}}$ are the crystalline pseudoangular momenta of the magnon and phonon modes, respectively, defined modulo $n$. Related pseudoangular-momentum conservation laws have been formulated for electron--photon and electron--phonon scattering processes~\cite{PhysRevB.97.195444}. An analogous rotational analysis of a magnetoelastic interaction $\mathcal{H}_{\mathrm{mp}}(\mathbf{k})$ preserving $C_n$ then constrains the coupling matrix element as
\begin{equation}\label{eq2}
\mathcal{G}_{\mu\nu}(\mathbf{k})\equiv
\bra{\mathrm{m}_{\mu\mathbf{k}}}
{\mathcal{H}}_{\mathrm{mp}}(\mathbf{k})
\ket{\mathrm{ph}_{\nu\mathbf{k}}}
=e^{i\frac{2\pi}{n}(l_{\mu}^{\mathrm{m}}-l_{\nu}^{\mathrm{ph}})}
\mathcal{G}_{\mu\nu}(\mathbf{k}).
\end{equation}
It follows that $\mathcal{G}_{\mu\nu}(\mathbf{k})\neq 0$ only if $l_{\mu}^{\mathrm{m}}=l_{\nu}^{\mathrm{ph}}\pmod n$. Thus, a $C_{n}$-symmetric magnetoelastic interaction permits hybridization only between magnon and phonon modes belonging to the same rotation sector.
\par
\begin{figure}[htbp]
\centering
\includegraphics[width=1\linewidth]{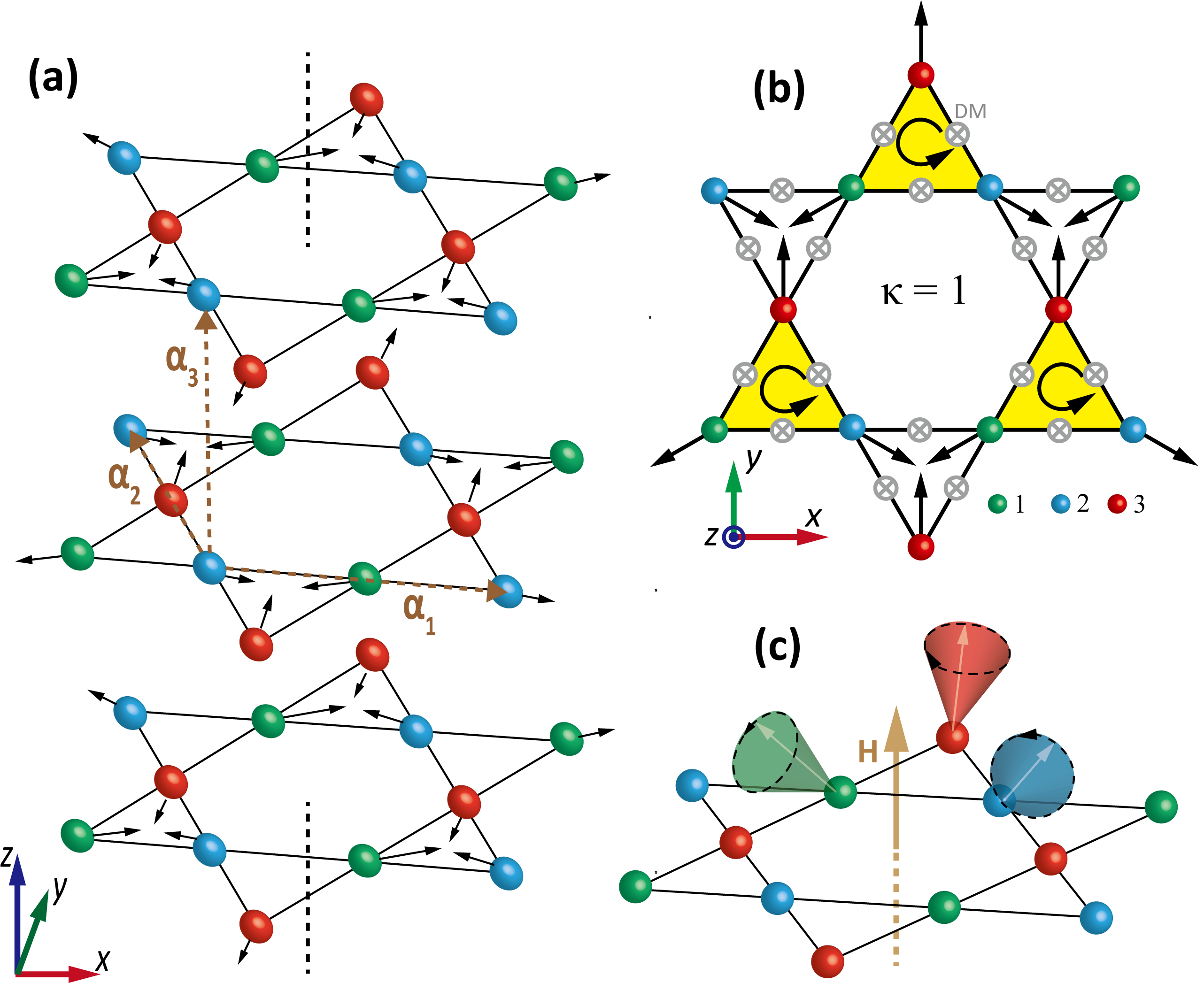}
\caption{Model geometry and magnetic configurations. (a) Stacked kagome lattice and its zero-field spin configuration. The dashed arrows denote the primitive lattice vectors $\boldsymbol{\alpha}_{1}$, $\boldsymbol{\alpha}_2$, and $\boldsymbol{\alpha}_3$. (b) Top view of a single layer with spin vector chirality $\kappa=1$. Gray $\otimes$ symbols denote DM vectors pointing into $-\hat{\mathbf z}$, while the curved arrows specify the oriented-bond convention; colors label sublattices 1--3. (c) Field-canted configuration for $\mathbf{H}\parallel\hat{\mathbf{z}}$, with the cones illustrating local spin precession.}
\label{fig1}
\end{figure}
\par
\textit{Model.}\textemdash To examine this selection rule, we consider the stacked kagome-lattice model shown in Fig.~\ref{fig1}(a), which combines noncollinear multisublattice order with a $C_3$-invariant $\mathrm{K}$ point. The magnetic Hamiltonian contains an intralayer nearest-neighbor antiferromagnetic exchange $J$, an out-of-plane DM interaction $D$, an antiferromagnetic interlayer exchange $J_{c}$, and a Zeeman field $H$ along the stacking direction. At $H=0$, the out-of-plane DM interaction stabilizes a coplanar $120^\circ$ spin texture within each layer, while $J_{c}>0$ favors antiferromagnetic alignment between adjacent layers~\cite{PhysRevB.97.094412}. Although this stacking nominally doubles the magnetic unit cell, the spin textures in neighboring layers are related by a uniform $\pi$ spin rotation about the $z$ axis. A one-layer translation combined with this spin rotation therefore maps the ordered state onto itself, allowing the spin-wave problem to be formulated in the three-sublattice kagome basis shown in Fig.~\ref{fig1}(b). Expanding the spin Hamiltonian about the field-canted spin texture to quadratic order in Holstein--Primakoff bosons~\cite{Holstein1940} yields
\begin{equation}
\mathcal{H}_{\mathrm{m}}=\frac{1}{2}\sum_{\mathbf{k}}
\psi^{\dagger}_{m\mathbf{k}}
\mathrm{H}_{\mathrm m}(\mathbf{k})
\psi_{m\mathbf{k}},
\end{equation}
where 
\begin{equation}
\psi_{m\mathbf{k}}=
\begin{pmatrix}
\hat{\mathbf{a}}_{\mathbf{k}}\\
\hat{\mathbf{a}}^{\dagger}_{-\mathbf{k}}
\end{pmatrix},
\qquad
\hat{\mathbf{a}}_{\mathbf{k}}=
\begin{pmatrix}
\hat{a}_{1\mathbf{k}}\\
\hat{a}_{2\mathbf{k}}\\
\hat{a}_{3\mathbf{k}}\\
\end{pmatrix},
\end{equation}
with 1--3 labeling the three kagome sublattices. The local-frame transformation and explicit form of $\mathrm{H}_{m}(\mathbf{k})$ are given in the Supplemental Material.
\par
\begin{figure*}[htbp]
\centering
\includegraphics[width=0.94\linewidth]{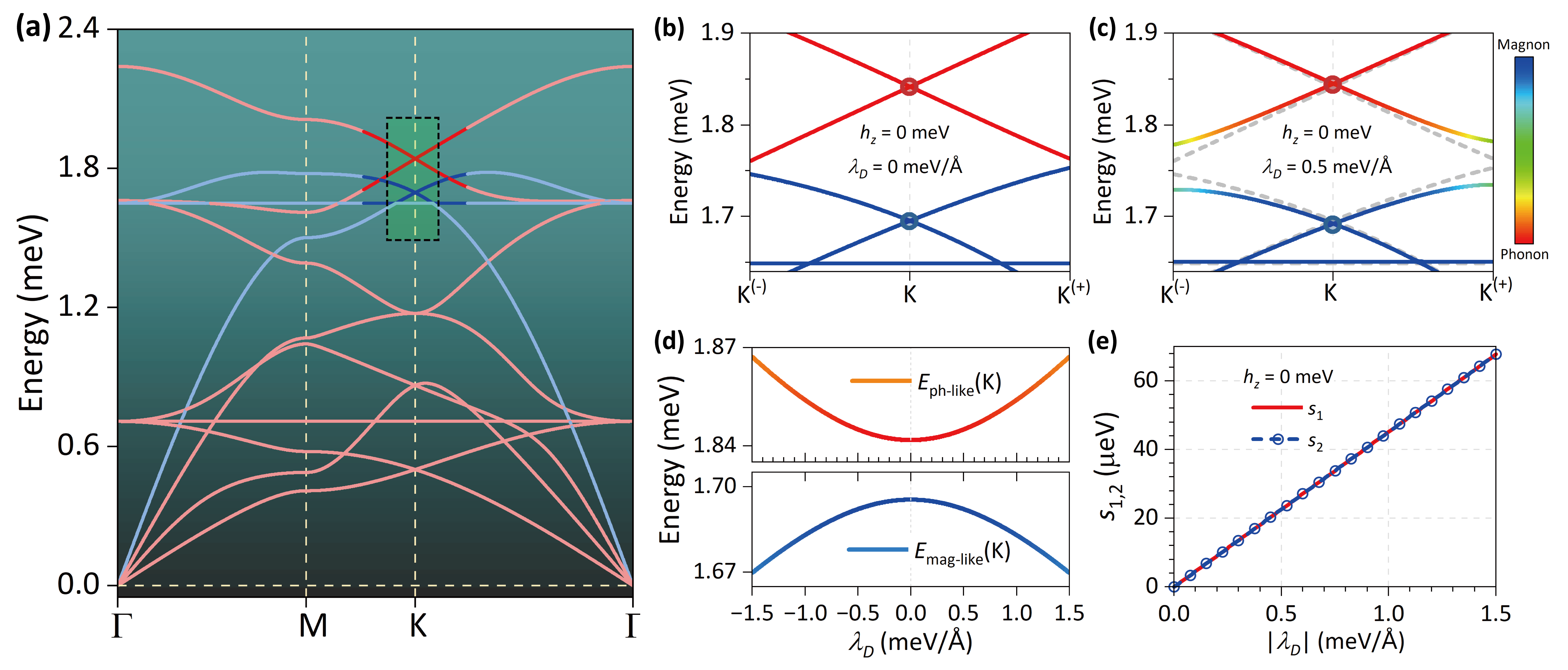}
\caption{Zero-field spectra and equivalent magnetoelastic coupling channels. (a) Uncoupled magnon (blue) and phonon (red) dispersions along $\Gamma$--$\mathrm{M}$--$\mathrm{K}$--$\Gamma$. The dashed box marks the $\mathrm{K}$-point region of interest. (b) Enlarged spectrum near $\mathrm{K}$ for $\lambda_D=0$, with $\mathrm{K}^{(-)}$ and $\mathrm{K}^{(+)}$ denoting momenta on either side of $\mathrm{K}$. (c) Corresponding spectrum for $\lambda_D=0.5~\mathrm{meV/\mathring{A}}$. Solid bands are colored by magnon weight, from phonon-like (red) to magnon-like (blue), and gray dashed lines denote the uncoupled dispersions. (d) Energies of the magnon-like and phonon-like doublets at $\mathrm{K}$ versus $\lambda_D$; both remain twofold degenerate. (e) Singular values $s_1$ and $s_2$ of the magnon--phonon coupling matrix between the two doublets at $\mathrm{K}$ versus $\lvert\lambda_D\rvert$, showing two equivalent orthogonal coupling channels.}
\label{fig2}
\end{figure*}
\par
The lattice dynamics is then described by a harmonic force-constant model in the same three-sublattice unit cell. To place the phonon sector on the same energy footing as the magnon Hamiltonian, we introduce dimensionless canonical coordinates,
\begin{equation}
\mathcal{H}_{\mathrm{ph}}=
\frac{1}{2}\sum_{\mathbf k}
\left[
E_{0}\mathbf{P}^{T}_{-\mathbf k}\mathbf{P}_{\mathbf k}+
\frac{1}{E_{0}}\mathbf{U}^{T}_{-\mathbf k}\mathcal{D}(\mathbf k)\mathbf{U}_{\mathbf k}
\right],
\end{equation}
where $\mathbf{U}_{\mathbf{k}}$ and $\mathbf{P}_{\mathbf{k}}$ are the dimensionless canonical displacement and momentum coordinates, respectively, $E_{0}$ is an auxiliary reference energy introduced by the rescaling, and $\mathcal{D}(\mathbf{k})=\hbar^2\Phi(\mathbf k)/M$ is the dynamical matrix expressed in energy-squared units. The explicit definitions of the scaled coordinates and the construction of $\Phi(\mathbf k)$ are given in the Supplemental Material.
\par
We subsequently couple the two parts by allowing phonon-induced bond distortions to modulate the intralayer magnetic interactions. Although exchange striction also generates a symmetry-allowed linear magnon--phonon coupling in a noncollinear magnet~\cite{PhysRevB.99.174435}, we focus here on the bond-length modulation of the out-of-plane DM interaction~\cite{PhysRevLett.123.167202,PhysRevB.104.045139,PhysRevB.105.L100402,bao2023direct}. This coupling directly links bond distortions to the vector spin chirality and provides a minimal setting for demonstrating the selection rule.
\par
For an intralayer nearest-neighbor bond $i\to j$, the instantaneous bond length is $r_{ij}=\lvert \mathbf{R}_{ij}^{(0)}+\mathbf{u}_{j}-\mathbf{u}_{i} \rvert$, where $\mathbf{R}_{ij}^{(0)}=\mathbf{R}_{j}^{(0)}-\mathbf{R}_{i}^{(0)}$ is the equilibrium bond vector and $\mathbf{u}_{i}$ denotes the physical displacement of site $i$. Expanding the bond-length-dependent DM coupling $D(r_{ij})$ to linear order about $r_{0}=\lvert \mathbf{R}_{ij}^{(0)}\rvert$ gives
\begin{equation}
D(r_{ij})\simeq D+\lambda_{D}\hat{\mathbf{e}}_{ij}\cdot(\mathbf{u}_{j}-\mathbf{u}_{i}),
\end{equation}
where $D=D(r_{0})$, $\lambda_{D}=(\partial D/\partial r)_{r_{0}}$, and $\hat{\mathbf{e}}_{ij}=\mathbf{R}^{(0)}_{ij}/r_{0}$. The equilibrium contribution is already included in $\mathcal{H}_{\mathrm{m}}$, while the displacement-linear term yields
\begin{equation}\label{eq7}
\mathcal{H}_{\mathrm{mp}} = \lambda_{D}\sum_{\left<ij\right>_{\mathrm{NN}}}\eta_{ij}
\left[
\hat{\mathbf{e}}_{ij}\cdot(\mathbf{u}_{j}-\mathbf{u}_{i})
\right]
\hat{\mathbf{z}}\cdot(\mathbf{S}_{i}\times \mathbf{S}_{j}).
\end{equation}
The sum runs over each physical intralayer nearest-neighbor bond once with a fixed orientation, and $\eta_{ij}=-\eta_{ji}=\pm 1$ specifies the directed DM convention through $\mathbf{D}_{ij}=\eta_{ij}D\hat{\mathbf{z}}$. For the bond arrows in Fig.~\ref{fig1}(b), we choose $\eta_{ij}=-1$. 
\par
Transforming the spins to their local frames and retaining terms linear in both the lattice displacement and Holstein--Primakoff bosons converts Eq.~\eqref{eq7} into a bilinear magnon--phonon coupling,
\begin{equation}
\mathcal{H}_{\mathrm{mp}}=\frac{1}{2}\sum_{\mathbf{k}}
\left[
\mathbf{U}_{-\mathbf{k}}^{T}\mathrm{V}_{\mathrm{Um}}(\mathbf{k})\psi_{m\mathbf{k}}+
\psi^{\dagger}_{m\mathbf{k}}\mathrm{V}^{\dagger}_{\mathrm{Um}}(\mathbf{k})\mathbf{U}_{\mathbf{k}}
\right],
\end{equation}
which defines the displacement--magnon coupling matrix $\mathrm{V}_{\mathrm{Um}}(\mathbf{k})$. Collecting the three quadratic contributions, we write
\begin{equation}
\mathcal{H}=\frac{1}{2}\sum_{\mathbf{k}}
\Psi_{\mathbf{k}}^{\dagger}
\mathrm{H}_{\mathrm{tot}}(\mathbf{k})
\Psi_{\mathbf{k}},
\end{equation}
in the composite basis $\Psi_{\mathbf{k}}=\left(\psi^{T}_{m\mathbf{k}},\mathbf{P}^{T}_{\mathbf{k}},\mathbf{U}^{T}_{\mathbf{k}}\right)^{T}$, with
\begin{equation}
\mathrm{H}_{\mathrm{tot}}(\mathbf{k})=
\begin{pmatrix}
\mathrm{H}_{\mathrm m}(\mathbf{k})&0&\mathrm{V}^{\dagger}_{\mathrm{Um}}(\mathbf{k})\\
0&E_{0}\mathbb{I}_{9}&0\\
\mathrm{V}_{\mathrm{Um}}(\mathbf{k})&0&\mathcal{D}(\mathbf{k})/E_{0}
\end{pmatrix}.
\end{equation}
Solving the corresponding bosonic generalized eigenvalue problem yields the positive-energy, positive-norm magnon-polaron branches~\cite{COLPA1978327}. Projecting $\mathrm{V}_{\mathrm{Um}}(\mathbf{k})$ onto the normalized bare magnon and phonon modes yields the mode-resolved couplings $\mathcal{G}_{\mu\nu}(\mathbf{K})$ introduced in Eq.~\eqref{eq2}, thereby connecting the microscopic magnetoelastic interaction directly to the pseudoangular-momentum selection rule. Details of the matrix construction and bosonic normalization are given in the Supplemental Material.
\par
\textit{Zero-field magnon--phonon coupling.}\textemdash We first examine the magnon and phonon spectra at zero field, $h_{z}=0$, where the ordered moments remain coplanar. Here, $h_z$ denotes the Zeeman field in energy units (see Supplemental Material). The uncoupled spectrum exhibits a twofold-degenerate magnon pair and a twofold-degenerate phonon pair at the $\mathrm{K}$ point, separated by a finite energy detuning [Figs.~\ref{fig2}(a) and \ref{fig2}(b)]. For finite $\lambda_D$, the two doublets hybridize, as evidenced by spectral-weight transfer and level repulsion [Fig.~\ref{fig2}(c)]. Nevertheless, the hybridization remains nonselective: the two members of each $\mathrm{K}$-point doublet undergo identical energy renormalization, preserving both degeneracies throughout the range of $\lambda_D$ considered [Fig.~\ref{fig2}(d)]. Thus, finite magnetoelastic coupling alone does not generate mode selectivity at zero field.
\par
To quantify the coupling in a basis-independent manner, we project the magnetoelastic interaction onto the magnon and phonon doublets at $\mathrm{K}$. The mode-resolved matrix elements defined in Eq.~\eqref{eq2}, with $\mu,\nu=1,2$, form a $2\times 2$ coupling matrix $\mathrm{G}(\mathrm{K})$, with $[\mathrm{G}(\mathrm{K})]_{\mu\nu}=\mathcal{G}_{\mu\nu}(\mathrm{K})$. Because the basis within either degenerate subspace is arbitrary, the individual matrix elements are basis dependent. We therefore characterize the coupling by the singular values $s_1$ and $s_2$ of $\mathrm{G}(\mathrm{K})$, which measure the basis-independent strengths of the two orthogonal coupling channels. As shown in Fig.~\ref{fig2}(e), both singular values scale linearly with $\lvert\lambda_D\rvert$, consistent with the displacement-linear form of $\mathcal{H}_{\mathrm{mp}}$, and remain identical throughout the range considered. Thus, the equality $s_1=s_2$ demonstrates that the two orthogonal channels are equally coupled, with neither favored at zero field.
\par
\begin{figure}[htbp]
\centering
\includegraphics[width=0.98\linewidth]{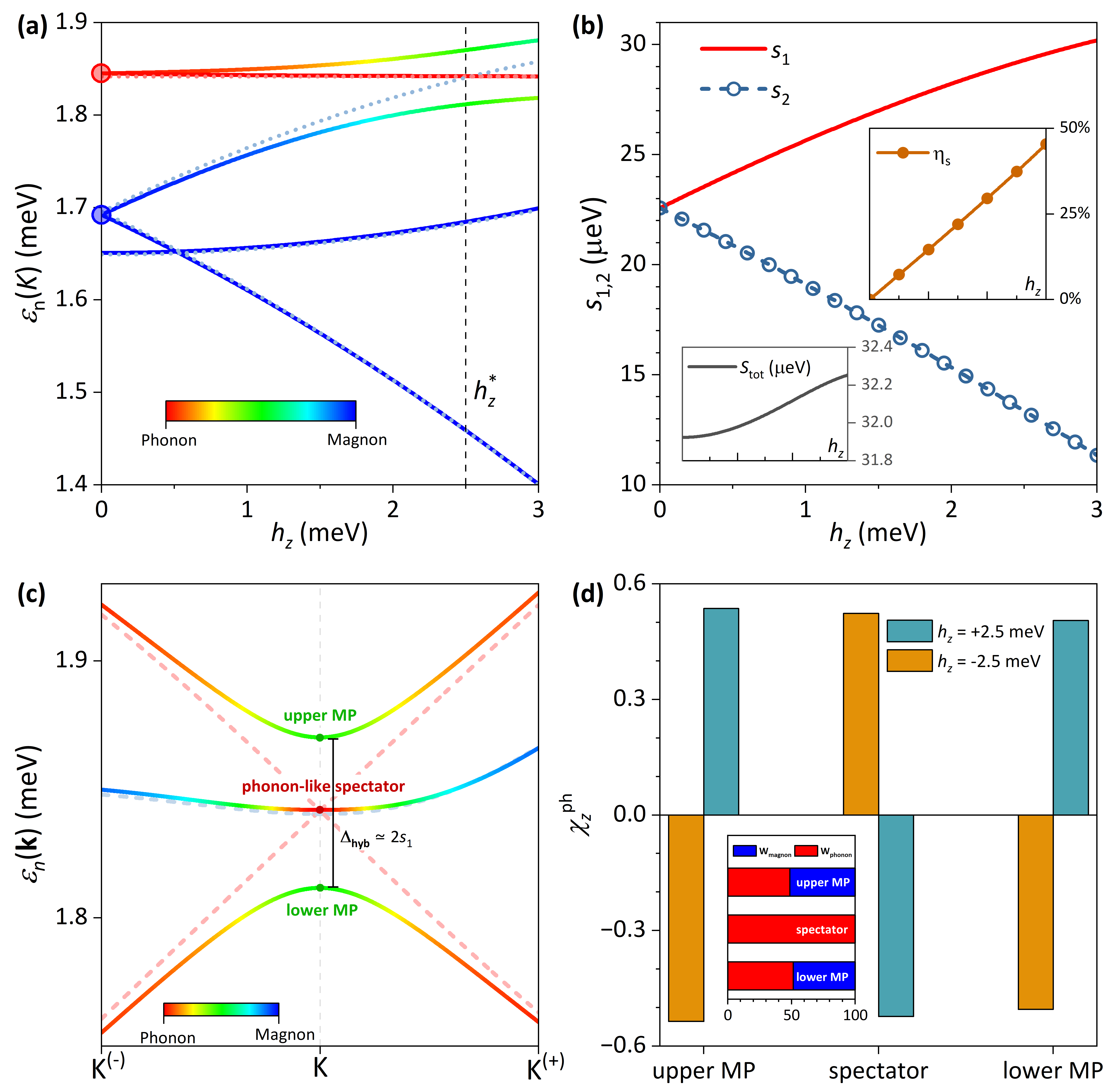}
\caption{Field-induced mode selectivity at $\mathrm{K}$ for $\lambda_D=0.5~\mathrm{meV/\mathring{A}}$. (a) Hybrid-mode energies at $\mathrm{K}$ versus $h_z$, colored by magnon weight; pale dashed lines denote the uncoupled energies. Open circles mark the zero-field doublets, and the vertical dashed line indicates $h_z^*=2.5~\mathrm{meV}$. (b) Singular values $s_1$ and $s_2$ of the projected magnon--phonon coupling matrix versus $h_z$. Insets show the total coupling strength $s_{\mathrm{tot}}=\sqrt{s_1^2+s_2^2}$ and channel imbalance $\eta_s=\lvert s_1-s_2\rvert/(s_1+s_2)$. (c) Hybrid spectrum near $\mathrm{K}$ at $h_z^*$, showing the upper and lower magnon-polaron (MP) branches and the phonon-like spectator mode; dashed lines denote the uncoupled dispersions. (d) Rotational polarization $\chi_z^{\mathrm{ph}}$ of the phonon components of the three hybrid branches for $h_z=\pm h_z^*$. The inset shows their corresponding magnon and phonon weights.}
\label{fig3}
\end{figure}

\textit{Field-induced mode selectivity.}\textemdash We next turn to finite $h_z$, which cants the ordered moments out of the kagome plane [Fig.~\ref{fig1}(c)] and lifts the equivalence of the two zero-field coupling channels. As shown in Fig.~\ref{fig3}(a), the field removes the $\mathrm{K}$-point magnon degeneracy and shifts the two branches in opposite directions, while the bare phonon doublet remains degenerate. One magnon branch is thereby tuned into energy resonance with the phonons at $h_z^*=2.5~\mathrm{meV}$. At finite field, we evaluate the same two channel strengths by retaining the complete positive-energy magnon subspace in the projection (see Supplemental Material). The field redistributes the magnetoelastic coupling between the two channels: $s_1$ increases while $s_2$ decreases, whereas their total strength $s_{\mathrm{tot}}$ varies only weakly [Fig.~\ref{fig3}(b)]. Thus, the zero-field equality $s_1=s_2$ evolves into a pronounced channel imbalance, signaling the field-induced inequivalence of the two symmetry sectors.
\par
At $h_z=h_z^*$, the field-tuned magnon is resonant with both degenerate phonons, yet hybridizes strongly with only one of them. The resulting upper and lower magnon-polaron (MP) branches are split by $\Delta_{\mathrm{hyb}}\simeq 2s_1$, while the orthogonal phonon combination remains essentially unmixed as a phonon-like spectator [Fig.~\ref{fig3}(c)]. The simultaneous presence of a single avoided crossing and a spectator mode therefore demonstrates that energy resonance alone is insufficient to determine hybridization.
\par
To characterize the lattice rotation of the participating phonon channel, we define the normalized out-of-plane rotational polarization of its phonon component as $\chi_z^{\mathrm{ph}}=2\sum_j\mathrm{Im}
\left[(u_{jx}^{\mathrm{ph}})^{*}u_{jy}^{\mathrm{ph}}\right]/
\sum_j\left(|u_{jx}^{\mathrm{ph}}|^2+|u_{jy}^{\mathrm{ph}}|^2\right)$, whose sign specifies the sense of ionic rotation. At $h_z=+h_z^*$, the phonon components of the two MP branches carry the same rotational polarization, whereas the spectator carries the opposite one [Fig.~\ref{fig3}(d)]. The mode weights in the inset show that the MP branches are nearly equal magnon--phonon mixtures at resonance, while the spectator remains almost purely phononic. Reversing the field switches the selected rotational polarization, so that the opposite phonon channel forms the MP pair. Although $\chi_z^{\mathrm{ph}}$ characterizes the rotational nature of the selected mode, it does not by itself determine the selection rule; the underlying criterion is encoded in the crystalline phase structure of the modes, as shown below.
\par
\par
\begin{figure*}[htbp]
\centering
\includegraphics[width=0.94\linewidth]{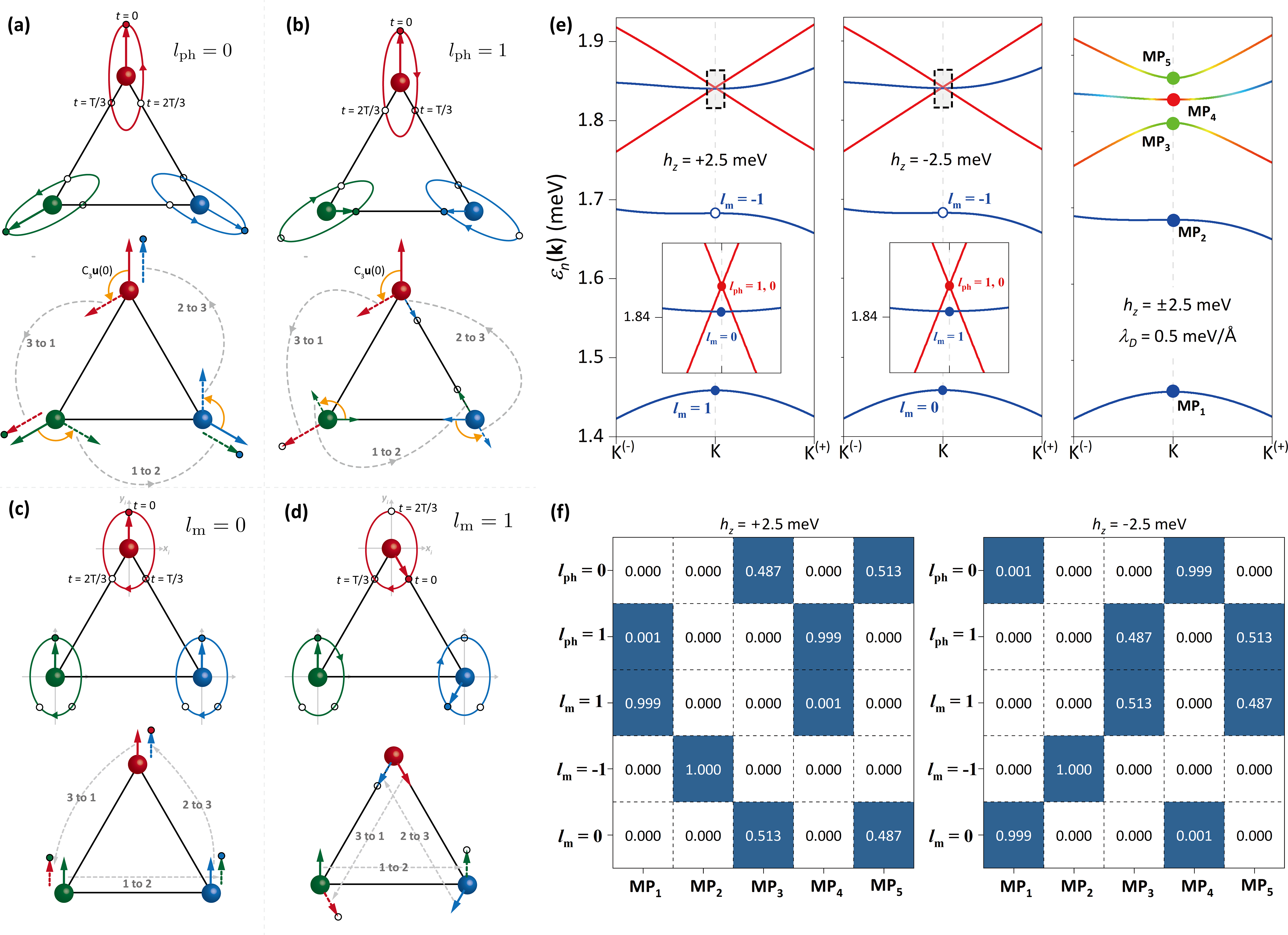}
\caption{Crystalline pseudoangular momentum and hybrid-mode decomposition at K. (a),(b) Displacement patterns and $C_3$ transformations of the phonon modes with $l_{\mathrm{ph}}=0$ and 1, respectively. (c),(d) Spin-precession patterns and $C_3$ transformations of the magnon modes with $l_{\mathrm{m}}=0$ and 1, respectively, represented in the local spin frames. (e) Uncoupled spectra for $h_z=\pm2.5$ meV (left and middle) and the coupled spectrum for $\lambda_D = 0.5~\mathrm{meV/\mathring{A}}$ (right). (f) Relative projection weights of the hybrid modes onto the selected uncoupled magnon and phonon modes for the two field directions. Filled cells denote finite mode participation.}
\label{fig4}
\end{figure*}
\par
\textit{Pseudoangular-momentum matching.}\textemdash To reveal the crystalline phase structure underlying the field-induced selectivity, we determine the pseudoangular momentum directly from the real-space action of $C_3$. For a mode with time dependence $e^{-i\omega t}$, the rotation eigenvalue $e^{-i2\pi l/3}$ in Eq.~\eqref{eq1} implies that its $C_3$-transformed configuration at $t=0$ coincides with the original mode after a time evolution $lT/3$, where $T=2\pi/\omega$. The upper parts of Figs.~\ref{fig4}(a) and \ref{fig4}(b) show the ionic trajectories and relative phases of the two phonon modes, while the lower parts show the corresponding configurations after applying $C_3$. Because the phonon eigenvectors are expressed in global Cartesian coordinates, $C_3$ rotates the displacement vectors while cyclically permuting the sublattices, including the associated Bloch phases. Comparing the transformed and time-evolved configurations identifies the two modes as $l_{\mathrm{ph}}=0$ and $1$, respectively.
\par
Figures~\ref{fig4}(c) and \ref{fig4}(d) apply the same construction to the magnon modes. Only the local spin-precession planes are shown, with the marked positions indicating the relative phases of the transverse fluctuations. Because the magnon amplitudes are defined in local spin frames, the spin rotation associated with $C_3$ is already incorporated into the local-frame transformation. The rotation representation therefore acts through cyclic sublattice permutation and the corresponding Bloch phases, without an additional rotation of the displayed spin vectors. This identifies the two magnon modes as $l_{m}=0$ and $1$, the same two pseudoangular-momentum sectors carried by the phonon doublet.
\par
We next label all uncoupled modes in the relevant energy window by their pseudoangular momenta [left and middle panels of Fig.~\ref{fig4}(e)]. For completeness, we include the third magnon branch lying between the two field-shifted branches, which carries $l_{\mathrm{m}}=-1$. Since the lattice sector is unaffected by the field, the phonon doublet retains $l_{\mathrm{ph}}=0$ and $1$. The magnetic sector, however, behaves differently: upon $h_{z}\to -h_{z}$, the two field-shifted magnon branches interchange their $l_{\mathrm{m}}=0$ and 1 assignments, while the intermediate $l_{\mathrm{m}}=-1$ branch remains unchanged. As a result, the magnon brought into resonance with the phonon doublet changes from the $l_{m}=0$ sector at $+h_{z}^{*}$ to the $l_{\mathrm{m}}=1$ sector at $-h_{z}^{*}$.
\par
Despite this interchange, the two field directions yield the same set of hybrid-mode energies [right panel of Fig.~\ref{fig4}(e)]. This apparent spectral equivalence conceals a field-reversed reorganization of the underlying rotation-sector content. To uncover this hidden change, we decompose the five coupled modes $\mathrm{MP}_{1-5}$ into the uncoupled magnon and phonon basis. As shown in Fig.~\ref{fig4}(f), at $+h_{z}^{*}$, the resonant $\mathrm{MP}_3$ and $\mathrm{MP}_5$ branches are nearly equal mixtures of the $l_{\mathrm{m}}=0$ magnon and the $l_{\mathrm{ph}}=0$ phonon. The off-resonant $l_{\mathrm{m}}=1$ magnon and the $l_{\mathrm{ph}}=1$ phonon remain concentrated in $\mathrm{MP}_1$ and $\mathrm{MP}_4$, respectively, while the $l_{\mathrm{m}}=-1$ magnon forms the unmixed $\mathrm{MP}_2$ mode. Reversing the field transfers the strong hybridization to the $l_{\mathrm{m}}=l_{\mathrm{ph}}=1$ sector, while the $l_{\mathrm{m}}=l_{\mathrm{ph}}=0$ sector becomes off resonant. Thus, although the hybrid-mode energies are unchanged, their underlying magnon and phonon constituents are interchanged. More generally, each hybrid mode contains bare components from only a single pseudoangular-momentum sector: the allowed components are fixed by crystalline symmetry, whereas their weights are controlled by energetic proximity.
\par
\textit{Summary.}\textemdash We have shown that crystalline pseudoangular momentum governs selective magnon–phonon hybridization in noncollinear multisublattice magnetic systems. In a stacked kagome model with noncollinear $120^\circ$ order, an out-of-plane field tunes a specific magnon rotation sector into resonance with the phonon doublet, while field reversal switches the participating phonon channel and reverses its rotational polarization. Remarkably, this reorganization of the underlying mode content occurs without changing the hybrid-mode energies. More broadly, our results separate two distinct ingredients of hybridization: rotation-sector compatibility determines which modes are allowed to mix, whereas energetic resonance controls the resulting degree of mixing. Local rotational polarization therefore characterizes the selected phonon component but does not, by itself, constitute the underlying selection criterion. Instead, crystalline pseudoangular momentum provides a symmetry-based handle for switching the internal rotation-sector composition of magnon polarons without necessarily altering their excitation energies. Such field-controlled selectivity provides a route toward symmetry-resolved control of magnon-polaron composition, with potential signatures in spectroscopic and transport measurements.
\begin{acknowledgments}
This work was funded by National Key R\&D Program of China (No.2022YFA1203100), National Natural Science Foundation of China (12504013), and Basic Research Program of Jiangsu (BK20250647).
\end{acknowledgments}
%\begin{thebibliography}{}
\normalem
\bibliography{ref}
%\end{thebibliography}
\end{document}